\documentclass[superscriptaddress,twocolumn,showpacs,prl,floatfix]{revtex4}

\usepackage{color}
\usepackage{tabularx}
\usepackage{epsfig}
\usepackage{amsmath}
\usepackage{graphicx}

\begin{document}

\title{$W$ : an alternative phenomenological coupling parameter for
  model systems}

\author{P. H. Lundow} \affiliation {Department of Theoretical Physics,
  KTH, SE-106 91 Stockholm, Sweden}

\author{I. A.~Campbell}
\affiliation{Laboratoire des Collo\"ides, Verres et Nanomat\'eriaux,
Universit\'e Montpellier II, 34095 Montpellier, France}

\begin{abstract}

We introduce a parameter $W(\beta,L)= (\pi\,\langle |m|
\rangle^2/\langle m^2 \rangle - 2)/(\pi-2)$ which like the kurtosis
(Binder cumulant) is a phenomenological coupling characteristic of the
shape of the distribution $p(m)$ of the order parameter $m$. To
demonstrate the use of the parameter we analyze extensive numerical
data obtained from density of states measurements on the canonical
simple cubic spin-$1/2$ Ising ferromagnet, for sizes $L=4$ to
$L=256$. Using the $W$-parameter accurate estimates are obtained for
the critical inverse temperature $\beta_c = 0.2216541(2)$, and for the
thermal exponent $\nu = 0.6308(4)$. In this system at least,
corrections to finite size scaling are significantly weaker for
the $W$-parameter than for the Binder cumulant.

\end{abstract}


\maketitle

\section{Introduction}
Studies of the critical properties of model systems using numerical
simulations are necessarily limited to samples of finite size. Finite
Size Scaling (FSS) techniques are essential in this context and a
Renormalization Group Theory (RGT) of FSS is well established
\cite{wegner:72, wegner:76, aharony:83, privman:84, privman:91,
  salas:00, pelissetto:02}.  At the critical point the shape of the
distribution of the order parameter $p(m)$ (throughout we will use
terminology appropriate to ferromagnetism) is independent of size $L$
to within finite size correction factors.

One widely used parameter characteristic of $p(m)$ is the kurtosis of
the distribution, $U_4 = \langle m^4 \rangle/\langle m^2 \rangle^2$.
The kurtosis is often expressed in terms of the Binder cumulant
\cite{binder:81}
\begin{equation}
g(\beta,L) = \frac{1}{2}\,\left(3-U_4(\beta,L)\right)
\label{gdef}
\end{equation}
because with this normalization $g(0,L) = 0$ in the high temperature
Gaussian limit (this is not strictly true for very small $L$) and
$g(\infty,L) = 1$ in the low temperature ferromagnetic (non-degenerate
ground state) limit ($\beta \equiv J/k_BT$ is as usual the normalized
inverse temperature).  For a given sample geometry (such as a
[hyper]cube) the thermodynamic (large $L$) limit of $g(\beta_c,L)$ is
a universal parameter for all systems in the same universality class.
Again in the large $L$ limit FSS theory shows that the slope $\partial
g(\beta,L)/\partial\beta \sim L^{1/\nu}$ at $\beta_c$ where $\nu$ is
the standard thermal critical exponent.  There are however corrections
to FSS which must be taken into account at all finite $L$.

The Binder parameter $g(\beta,L)$ is not the only distribution "shape"
parameter having these properties. We will introduce and illustrate on
the simple cubic $S=1/2$ Ising ferromagnet an alternative parameter
$W(\beta,L)$ which has some technical advantages at least in this
case, in particular having corrections to FSS which are weaker than
those of the Binder parameter. If this turns out to be a general
property the $W$-parameter could be very helpful for estimating
critical properties numerically in more difficult cases where
simulations are intrinsically restricted to more moderate size
samples. Already for the $3$d Ising ferromagnet we obtain rather
precise estimates for the critical inverse temperature $\beta_c$ and
the thermal critical exponent $\nu$ using this parameter.

\section{Phenomenological couplings}
A "phenomenological coupling" is broadly a parameter which becomes $L$
independent at $\beta_c$ in the thermodynamic limit
\cite{pelissetto:02}; as well as the Binder parameter the normalized
second moment correlation length $\xi(\beta,L)/L$ is an important
phenomenological coupling. It should be noticed that below $T_c$ both
the Binder parameter and $\xi(\beta,L)/L$ are defined in terms of
non-connected distribution sums. While $g(\beta,L)$ is defined such
that $g(\infty,L)=1$, the conventional definition of $\xi(\beta,L)$
leads to $\xi(\infty,L)/L = \infty$ for a system with a non-degenerate
ground state. A phenomenological coupling with a different
normalization, defined by $R_{\xi}(\beta,L) = \xi(\beta,L)/(L + \xi(\beta,L))$,
would be closer in spirit to the Binder cumulant.

The standard RGT FSS expression with leading correction terms for a
phenomenological coupling $R(\beta,L)$ such as $g(\beta,L)$, the
normalized correlation length $\xi/L(\beta,L)$, or the $W(\beta,L)$ to
be introduced below, is
\begin{equation}
R(\beta,L) = R(u_{\tau}L^{1/\nu}) +
v_{\omega}\,R_{\omega}(u_{\tau}L^{1/\nu})\,L^{-\omega} +
\cdots\nonumber
\end{equation}
\begin{equation}
\approx R_{c} + \left[\partial
R/\partial\tau\right]_{0}\,c_{\tau}\,\tau\, L^{1/\nu} +\cdots +
c_{\omega}\, L^{-\omega} + \cdots
\label{fss}
\end{equation}
where $\tau$ is the thermal scaling variable (for instance $\tau = 1 -
\beta/\beta_c$), $u_{\tau}$ is the thermal scaling field, $\omega =
\theta/\nu$ is a universal scaling correction exponent, and the other
parameters (critical temperature and critical amplitudes) are
non-universal constants appropriate for each particular system. The
second line is a good approximation as long as $\tau\, L^{1/\nu}\ll 1$.
Thus
\begin{equation}
R(\beta_c,L) = R_{c}\,\left(1 + c_{\omega}\,L^{-\omega} + \cdots\right) ,
\label{fssR}
\end{equation}
and
\begin{equation}
[\partial R/\partial\beta]_{\beta_c} = K\,L^{1/\nu}\,\left(1 +
K_{\omega}\,L^{-\omega} + \cdots\right)
\label{fssdRdb}
\end{equation}
Another important conclusion \cite{binder:81} is that the intersection
temperatures for $R(\beta,L)$ and $R(\beta,sL)$, denoted
$\beta_{\mathrm{cross}}(L,s)$, converge as
\begin{equation}
\beta_{\mathrm{cross}}-\beta_{c} \sim L^{-(\omega +1/\nu)}
\label{betacross}
\end{equation}

The parameter $W(\beta,L)$ which we introduce is a function of the
ratio of the variance of the modulus of $m$,
\begin{equation}
\chi_{\mathrm{mod}}=\left\langle(|m|-\langle|m|\rangle)^2\right\rangle
= \langle m^2 \rangle - \langle |m| \rangle ^2
\end{equation}
to the variance of $m$
\begin{equation}
\chi = \left\langle(m-\langle m \rangle)^2\right\rangle =
\langle m^2 \rangle - \langle m \rangle^2
\end{equation}
For a finite $L$ ferromagnet in zero applied field, which is the case
that we will discuss explicitly, the distribution $p(m)$ is always
symmetric so $\langle m \rangle = 0$ even below the critical
temperature, thus $\chi = \langle m^2 \rangle$

We will define the normalized parameter
%\begin{equation}
%W = 1-\pi(\chi_{mod}/\chi)/(\pi-2)
%\end{equation}
%or
%\begin{equation}
%W = 2-(4-\pi)[\langle m^2 \rangle/\langle |m|^2 \rangle]/(\pi-2)
%\label{Wdef}
%\end{equation}
\begin{equation}
  W = 1- \frac{\pi}{\pi-2}\,\frac{\chi_{\mathrm{mod}}}{\chi}
\end{equation}
or
\begin{equation}
  W = \frac{\pi\,U_2 - 2}{\pi - 2}
\end{equation}
where $U_2 = \langle|m|\rangle^2/\langle m^2 \rangle$.
The normalization has been chosen such that, as for the Binder
parameter, $W(\beta,L) = 0$ in the high temperature Gaussian limit and
$W(\beta,L) = 1$ in the low temperature ferromagnetic limit. As
$W(\beta,L)$ is also a parameter characteristic of the shape of the
distribution $p(m)$, it can be considered to be another
"phenomenological coupling" and so will share all the formal finite
size scaling properties of $g(\beta,L)$. 

By analogy with "kurtosis" (derived from the Greek word for a curve or
bulge) we propose to name $W$ the \lq\lq dichokurtosis\rq\rq,
referring to the process of dividing into two parts, i.e. a unimodal
distribution shifting into a bimodal one.

As a demonstration, extensive data on $W(\beta,L)$ will be discussed
for the canonical case of the simple cubic spin $S=1/2$ Ising
ferromagnet.  Though we will not discuss this point further here, the
properties of the distribution $p(|m|)$ are of particular interest
when the regime $T < T_c$ is studied as well as $T > T_c$. Above
$T_c$, $\langle m \rangle \equiv 0$ in zero applied field; the
connected and non-connected susceptibilities
\begin{equation}
\chi_{\mathrm{conn}} = \langle m^2 \rangle - \langle m_{h}^2 \rangle
\end{equation}
and
\begin{equation}
\chi_{\mathrm{non}} = \langle m^2 \rangle - \langle m_{h}^2 \rangle
\end{equation}
are identical. Below $T_c$ it is the connected susceptibility which is
physically significant in the thermodynamic limit. For finite $L$ the
distribution $p(m)$ consists approximately of two peaks centered on
$\pm \langle |m| \rangle$. As $\beta - \beta_c$ and $L$ increase this
approximation gets better and better because the peaks narrow so that
\begin{equation}
\langle |m| \rangle \approx \langle m_{h} \rangle
\end{equation}
where $\langle m_{h} \rangle$ is the magnetization that would be
measured in an infinitesimal applied field.  Hence
$\chi_{\mathrm{mod}}(\beta,L) \approx \chi_{\mathrm{conn}}(\beta,L)$.

\section{Numerical Methods}
The physical parameters for finite size samples from $L=4$ up to
$L=256$ ($16,777,216$ spins) were estimated using a density of states
function method. When studying a statistical mechanical model complete
information can in principle be obtained through the density of states
function. From complete knowledge of the density of states one can
immediately work with the microcanonical (fixed energy) ensemble and
of course also compute the partition function and through it have
access to the canonical (fixed temperature) ensemble as well. The main
problem here is that computing the exact density of states for systems
of even modest size is a very hard numerical task. However, several
sampling schemes have been given for obtaining approximate density of
states, of which the best known are the Wang-Landau \cite{wang:01} and
Wang-Swendsen \cite{wang:02} methods. In \cite{haggkvist:04} various
methods are discussed in a common framework.  For work in the
microcanonical ensemble the sampling methods give all the information
needed. Using them one can find the density of states in an energy
interval around the critical region and that is all that is required
for most investigations of the critical properties of the model.  It
should be noted that there is no standard technique for estimating the
error bars in the outputs of this class of method, other than
repeating the entire calculation a number of times which would be
extremely laborious.

For the present analysis a density of states function technique based
upon the same method as in \cite{haggkvist:07} was used though with
considerable numerical improvements for all $L$ studied here (adequate
improvements to the $L=512$ data set would unfortunately have been too
time-consuming). The microcanonical (energy dependent) data were
collected as described in \cite{haggkvist:04}. We use standard
Metropolis single spin-flip updates, sweeping through the lattice
system in a type-writer order. Measurements take place when the
expected number of spin-flips is at least the number of sites. For
high temperatures this usually means two sweeps between measurements
and four or five sweeps for the lower temperatures we used. Note that
in the immediate vicinity of $\beta_c$ the spin-flip probability is
very close to $50\%$.

For $L=256$, the largest lattice studied here, we have now amassed
between 500 and 3500 measurements on an interval of some 450000 energy
levels, where most samplings are near the critical energy.  For
$L=128$ we have between 5000 and 50000 measurements on some 150000
energy levels.  For $L\le 64$ the number of samplings is of course
vastly bigger.

Our measurements at each individual energy level include local energy
statistics and magnetisation moments.  The microcanonical data were
then converted into canonical (temperature dependent) data according
to the technique in \cite{lundow:09}. This gave us energy distributions
from which we may obtain energy cumulants (e.g. the specific heat) and
magnetization cumulants (e.g. the susceptibility).

Typically around 200 different temperatures were chosen to compute
these quantities, with a higher concentration near $\beta_c$
particularly for the larger $L$ so that one may use standard
interpolation techniques on the data to obtain intermediate
temperatures.  Magnetization distributions $p(m)(\beta,L)$ have also
been obtained for sizes from $L=4$ to $L=64$.

\section{Equilibration times}
We can make a critical comparison between the Binder parameter and the
$W$-parameter from the point of view of equilibration time.  At the
heart of the $W$-parameter is the ratio $U_2 = \langle
|m|\rangle^2/\langle m^2\rangle$, just as the ratio $U_4 = \langle m^4
\rangle / \langle m^2 \rangle^2$ is the basis for the
$g$-parameter. Since the $U_4$-ratio involves a fourth moment we
expect it to converge more slowly to its limit value than $U_2$, which
contains only a second moment.  We have measured the speed of
convergence by studying the respective variation coefficients
$\sigma/\mu$ as a function of the number of measurements $n$. As usual
$\sigma$ refers to the standard deviation of the measurements and
$\mu$ to the average measurement. This allows us to compare the two,
though the result will of course depend on the underlying
distribution. We have chosen to look at this for a simple cubic
lattice with $L=16$ at $\beta=0.225$, a temperature slightly below
where the distribution changes from unimodal to bimodal.

We perform $n$ measurements of $|m|$, $m^2$ and $m^4$ and take their
respective averages, giving us estimates of $\langle |m|\rangle$,
$\langle m^2\rangle$ and $\langle m^4\rangle$. The estimate of $U_2$
is now simply $\langle |m|\rangle^2/\langle m^2\rangle$ and for $U_4$
we use $\langle m^4\rangle/\langle m^2\rangle^2$. Repeating these
$n$-estimates a number of times (75 times for n=100000 and 75000 times
for n=100) gives us, in turn, an estimate of the variance $\sigma^2$
of the $U_2$- and $U_4$-estimates. As $\mu$ we use the average $U_2$-
and $U_4$-estimates.

\begin{figure}
\includegraphics[width=3.5in]{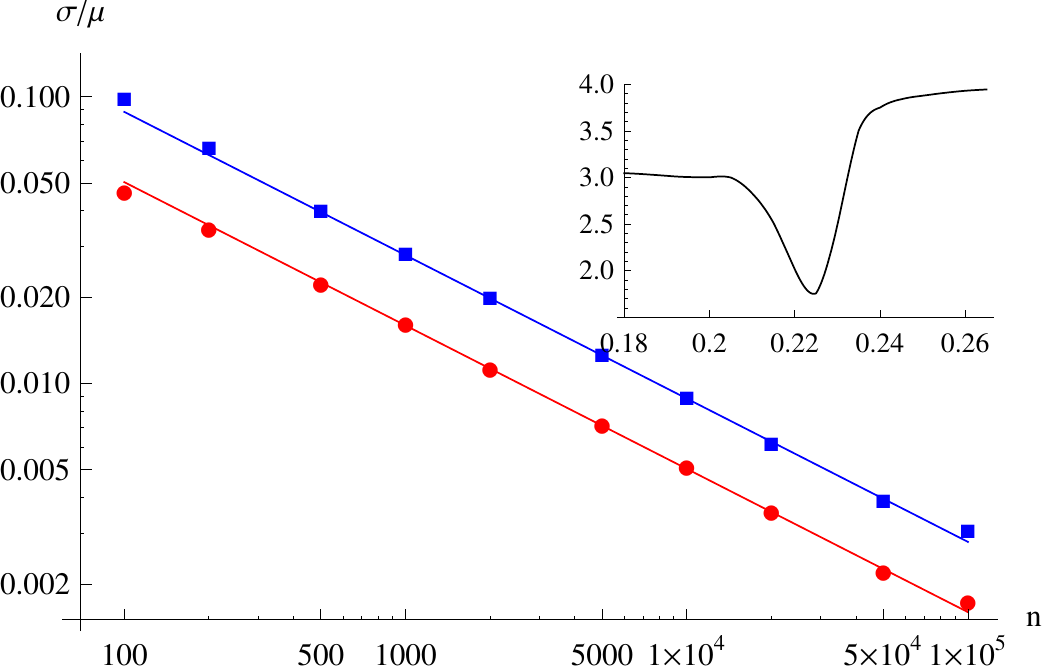}
\caption{(Color online) Variation coefficient $\sigma/\mu$ for $U_2$
  (red circles) and $U_4$ (blue squares) at $\beta=0.225$ for $L=16$
  plotted versus the number of measurements $n$ together with fitted
  lines with slope $-1/2$. The red line is $0.504/\sqrt{n}$ and blue
  line is $0.886/\sqrt{n}$. The inset shows the ratio between these
  two for a range of $\beta$, having the minimum $1.76$ at
  $\beta=0.225$.}  \protect\label{fig:1}
\end{figure}

In Figure~\ref{fig:1} we show $\sigma/\mu$ versus $n$ for $U_2$ and
$U_4$ for the 3d-lattice with $L=16$. We have fitted lines with slope
$-1/2$ since we expect the variation coefficient to decrease at the
rate $1/\sqrt{n}$.  We find that $\sigma/\mu$ scales as roughly
$0.504/\sqrt{n}$ for $U_2$ and $0.886/\sqrt{n}$ for $U_4$. 

Squaring the factor $0.886/0.504\approx 1.76$ gives that $U_4$
requires $1.76^2\approx 3.1$ times as many measurements as $U_2$ to
obtain the same statistical error $\sigma/\mu$ at $\beta=0.225$.

The factor $1.76$ is actually close to a worst case scenario for this
particular lattice.  For higher temperatures, i.e. $\beta<\beta_c$,
this factor takes a value close to three and for lower temperatures,
i.e. $\beta>\beta_c$, the factor quickly approaches a value close to
four, Figure~\ref{fig:1} inset. It also turns out that this worst case
factor actually increases with $L$. For $L=8$ we measured it to $1.74$
at $\beta=0.23$ while for $L=32$ we found it to be $1.83$ at
$\beta=0.2225$.

\begin{figure}
\includegraphics[width=3.5in]{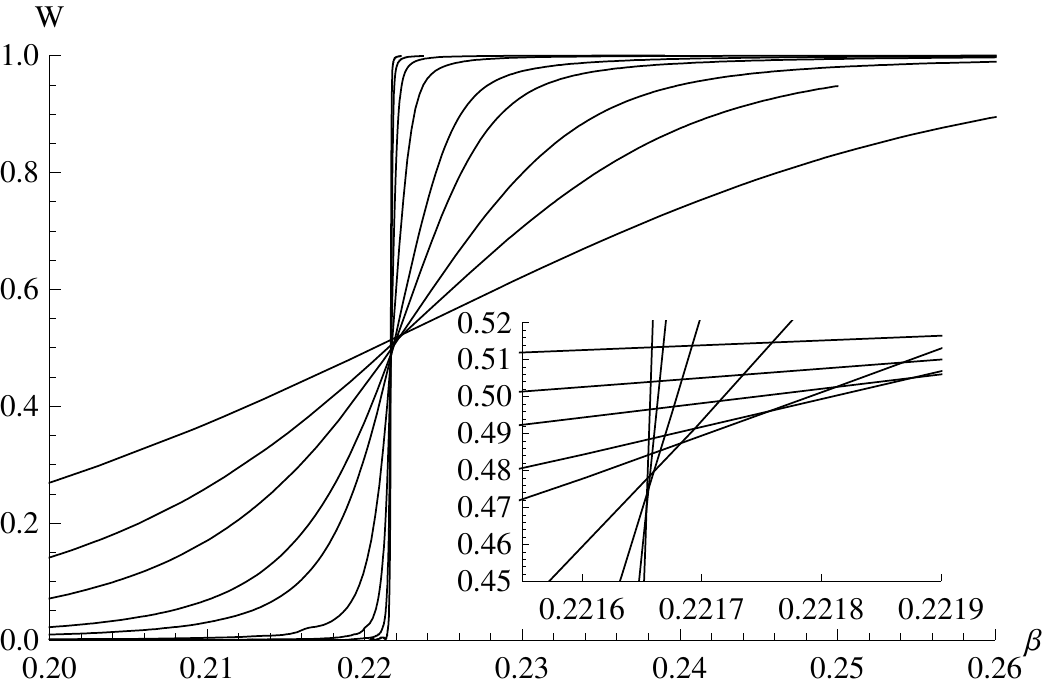}
\caption{(Color online) $W(\beta,L)$ versus $\beta$ for $L=4$
  (smallest slope) to $L=256$ (strongest slope). The inset shows a
  zoomed in picture near $\beta_c$.  } \protect\label{fig:2}
\end{figure}

\begin{figure}
\includegraphics[width=3.5in]{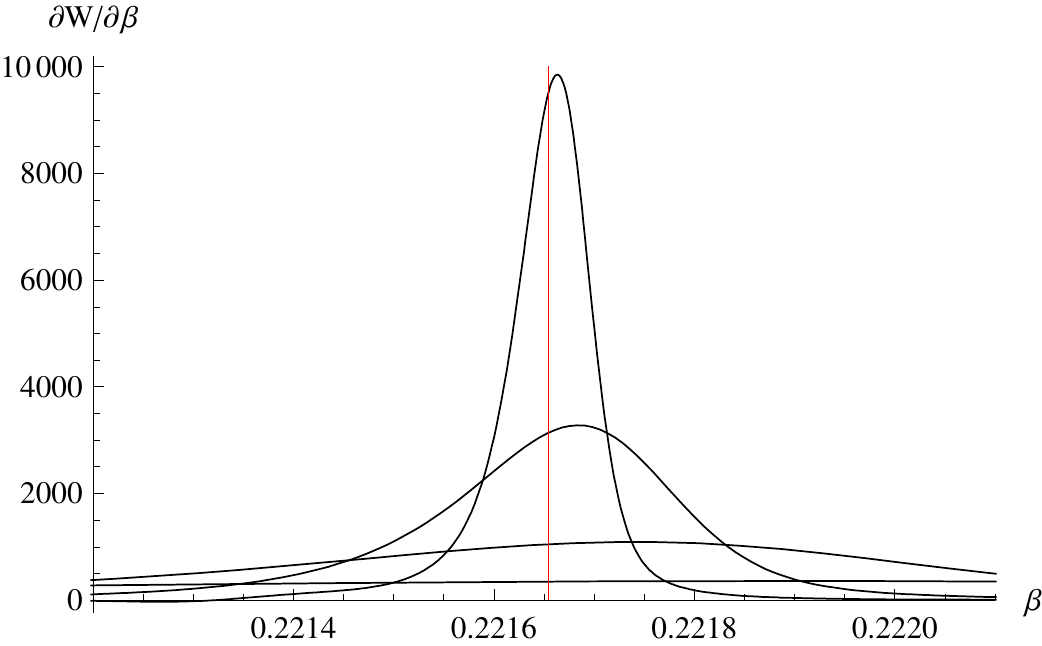}
\caption{(Color online) $\partial W/\partial\beta$ versus $\beta$ for
  $L=32,64,128,256$ where the maximum increases with $L$. The red
  vertical line is located at $\beta_c=0.2216541$.  }
\protect\label{fig:3}
\end{figure}

\begin{figure}
\includegraphics[width=3.5in]{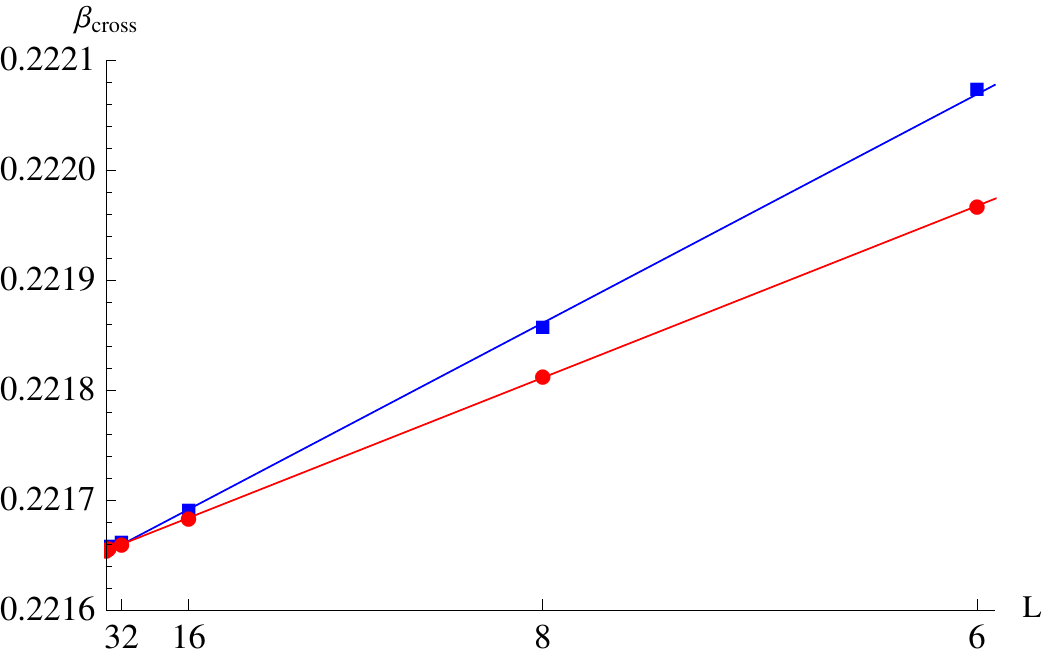}
\caption{(Color online) Crossing points $\beta_{\mathrm{cross}}$
  versus $1/L^{2.40}$ for $L=6,8,16,32,64,128$ for $W$ (red circles)
  and $g$ (blue squares) and fitted lines.}  \protect\label{fig:4}
\end{figure}

\begin{figure}
\includegraphics[width=3.5in]{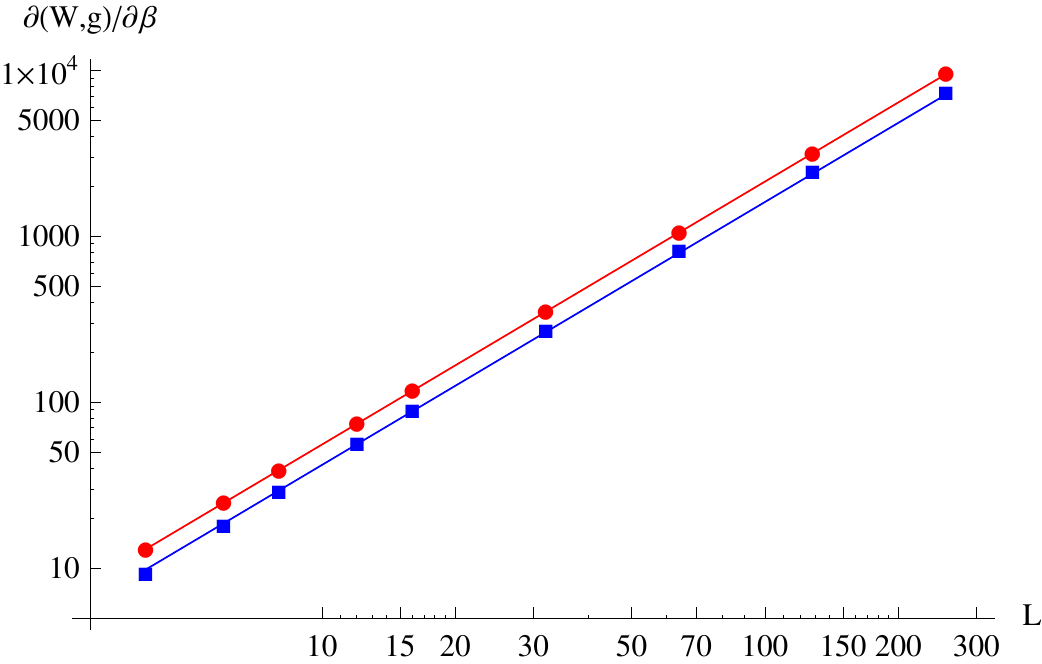}
\caption{(Color online) Log-log plot of $\partial W/\partial\beta$
  (red circles) and $\partial g/\partial\beta$ (blue squares) versus
  $L$ at $\beta=\beta_c=0.2216541$. Lines have slope $1/\nu$. }
\protect\label{fig:5}
\end{figure}

\begin{figure}
\includegraphics[width=3.5in]{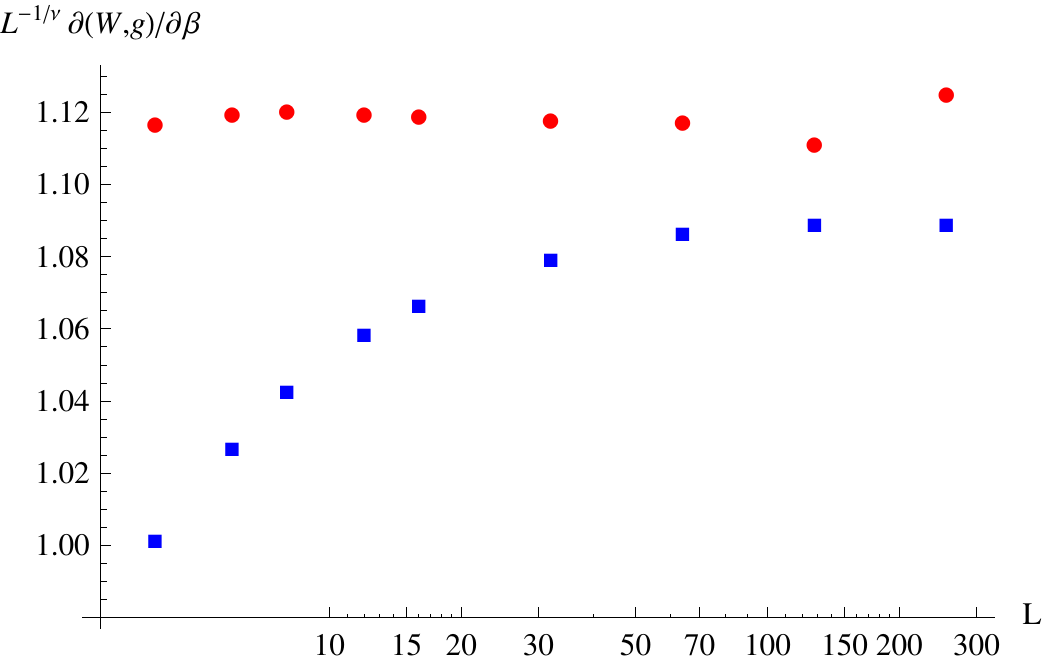}
\caption{(Color online) $L^{-1/\nu}\,\partial W/\partial\beta-0.32$
  (red circles) and $L^{-1/\nu}\,\partial g/\partial\beta$ (blue
  squares) at $\beta=\beta_c=0.2216541$ versus $L$. The $W$-points
  have been translated by $-0.32$ for easier comparison.}
\protect\label{fig:6}
\end{figure}

\begin{figure}
\includegraphics[width=3.5in]{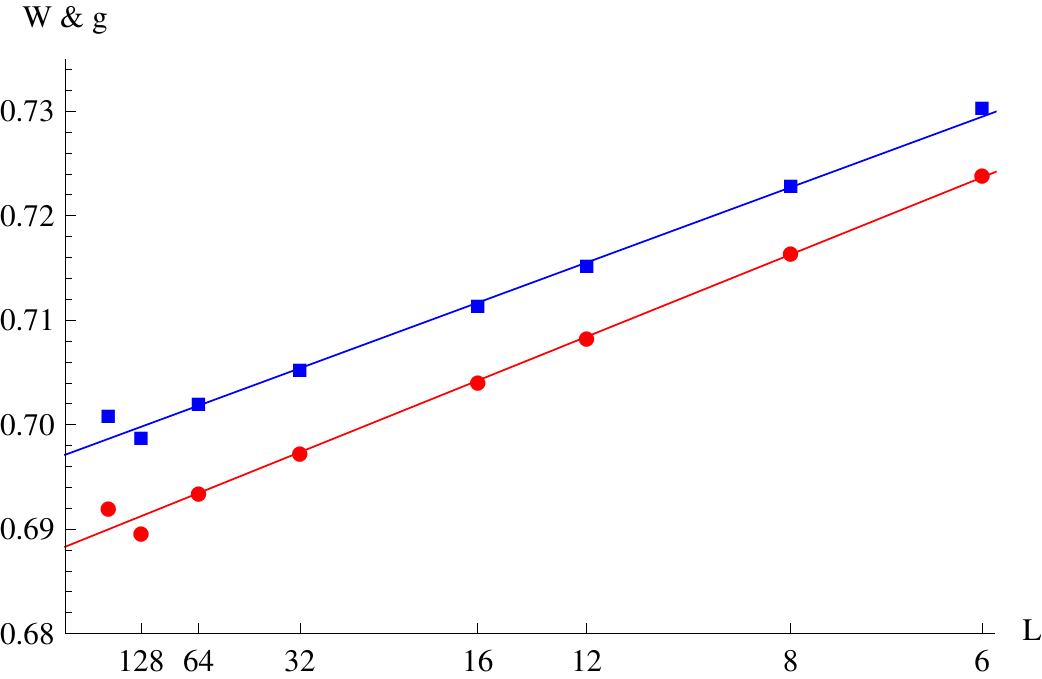}
\caption{(Color online) $W(\beta_c,L)+0.22$ (red circles) and
  $g(\beta_c,L)$ (blue squares) versus $1/L^{\omega}$ with
  $\omega=0.814$ for $L=6,8,16,32,64,128,256$. The $W$-points have
  been translated by $0.22$ for easier comparison.}
\protect\label{fig:7}
\end{figure}

\section{The 3d Ising Ferromagnet}
The spin-$1/2$ Ising ferromagnet on a simple cubic lattice is an
archetypical model system which has been very extensively
studied. Although there exist no exact results for any of the critical
parameters, $\beta_c$ and the critical exponents are known to high
precision thanks to RGT theory, high temperature series expansions
(HTSE), and numerical simulations (see
Refs. ~\cite{guida:97,butera:02,deng:03}). There is consensus that for
this system $\beta_c \approx 0.221655$, and for the universality class
$\nu \approx 0.630$, and $\omega \approx 0.81$. We will test our
$W$-data against these values shortly.

%Among the parameters registered in the $3$d Ising ferromagnet density
%of states calculations were $\langle m^2 \rangle$ and $\langle
%|m|\rangle$.
We show in Figure~\ref{fig:2} an overall view of the behavior of
$W(\beta,L)$.  It can be seen that on the scale of the figure the
curves $W(\beta,L)$ appear to intersect at a unique $L$-independent
inverse temperature which can obviously be identified with
$\beta_c$. The derivatives $\partial W/\partial\beta$ peak strongly at
$\beta_{W\!\max}(L)$ which at large $L$ approaches $\beta_c$, see
Figure~\ref{fig:3}.

%%%% New version
A blow-up of $W(\beta,L)$ in the critical region, Figure~\ref{fig:2}
inset, shows that there are finite size corrections leading to a weak
size dependence of the $[W(\beta,L),W(\beta,2L)]$ crossing points. A
plot of the intersection temperatures $\beta_{\mathrm{cross}}$ versus
$1/L^{\omega+1/\nu}$, where $\omega+1/\nu \approx 2.40$, is shown in
Figure~\ref{fig:4} for both $W$ and $g$ (the points for $L=64,128$ are
not visible due to the fast data collaps). Fitting a straight line to
the $W$-points for $L\ge 6$ versus $1/L^{\omega+1/\nu}$ gives
$\beta_c=0.2216541(5)$. We have here excluded the point $L=4$ since it
appears to deviate from the others in this case. The error estimate is
based on how the result depends on excluding a point from the fit and
on allowing the exponent $\omega+1/\nu$ to take different values
between $2.39$ and $2.41$. In fact, a best fit of the crossing points
for $L\ge 6$ to a simple formula $c_0+c_1\,L^{-\lambda}$ gives on
average, taken over fits after excluding one point, $c_0\approx
0.2216540(3)$ and $\lambda=2.40(3)$, where the error estimates
correspond to the standard deviation of the data set. Since
$\nu=0.630$ is known to a higher precision we therefore get
$\omega=0.814(30)$ which agrees with previous estimates.

Both of our $\beta_c$-estimates are consistent with the most precise
values from standard Monte-Carlo simulations $\beta_c = 0.22165452(8)$
\cite{deng:03}, $\beta_c=0.2216546(3)$ \cite{haggkvist:07}, and from
high temperature series analyses, $\beta_c = 0.221655(2)$
\cite{butera:02}.  However, the $g$-data seem to require more
correction to scaling than the $W$-data. If we want to fit a line to
the crossing points for $g$ versus $L^{\omega+1/\nu}$ then we need to
drop two more points ($L=6,8$) to get anything like this precision on
a $\beta_c$-estimate.

Henceforth setting $\beta_c=0.2216541$, let us proceed to investigate
the derivative data $[\partial W/\partial\beta]_{\beta_c}$ and
$[\partial g/\partial\beta]_{\beta_c}$ against $L$, which are shown as
log-log plots in Figure~\ref{fig:5}. The slopes should be equal to
$1/\nu$ in the large $L$ limit. It can be seen that both series of
points lie close to $1/0.63$ (slope of the lines).

Let us make a more demanding analysis of the slopes $1/\nu$ by fitting
lines to $k$-subsets of the points. Since we have $9$ data points,
i.e. we use $L\ge 4$, each $k$ then gives us $\binom{9}{k}$ different
slopes. If the data show any sign of inconsistency or a dependency on
$L$ then we expect this to show up in the form of different medians
and/or different slope intervals.  However, we get $\nu=0.6308$ for
$k=3,\ldots,9$, with the same value for both median and mean. The
quartile deviation of each slope set is about $0.0004$ for
$k=4,\ldots,7$. We therefore receive the estimate $\nu=0.6308(4)$. It
should be noted that only for the last three points of the $g$-data do
we receive a slope that agrees with this estimate.

An alternative way of locating $\beta_c$ is to locate the temperature
$\beta_c$ where the scaling of the derivatives depend least on
different $L$.  Choosing e.g. subsets of size $k=4$ the narrowest set
of slopes is obtained for $\beta_c=0.2216541$, give or take a step or
two in the last decimal. Since this agrees with our previous two
estimates of $\beta_c$ we can now give our final estimate of the
critical temperature as $\beta_c=0.2216541(2)$.

%% A more detailed fit to the $W$-data for $L=4$ to $L=256$ leads to
%% $\beta_c=0.2216541(1)$ and $\nu = 0.6308(4)$. This was obtained by
%% working with the logarithm of the derivatives $\partial
%% W/\partial\beta$ at some chosen $\beta$. We fitted a straight line to
%% each 4-subset of the 9 points thus providing us with 126 different
%% slopes (i.e. estimates of $1/\nu$). The minimum difference between the
%% maximum and minimum slopes was obtained at $\beta=0.2216541$, give or
%% take a step in the seventh decimal.  Thus we estimate
%% $\beta_c=0.2216541(1)$. At this $\beta_c$ the $\nu$ distribute between
%% $0.6301$ and $0.6318$ with a median of $0.6308$ and a quartile
%% deviation from the median of $0.0004$, thus giving us $\nu=0.6308(4)$.

Having established $\beta_c$ and $\nu$ we plot the derivatives of $W$
and $g$ in the more demanding form $[\partial W/\partial
  \beta]_{\beta_c}/L^{1/\nu}$ and $[\partial g/\partial
  \beta]_{\beta_c}/L^{1/\nu}$ in Figure~\ref{fig:6}.  The $g$-data
clearly show characteristic FSS corrections
\begin{equation}
g(\beta_c,L) = g(\beta_c,\infty)\,\left(1 + a_{\omega}\,L^{-\omega} +
\cdots\right)
\end{equation}
at small and moderate $L$ while the $W$-data show only weak and
apparently random scatter due to statistical errors, i.e. the
analogous correction term for $W(\beta_c,L)$ appears negligible within
the present precision. This means that to extract an estimate of $\nu$
a two parameter fit is sufficient for the $W$ derivative data while a
four parameter fit is needed for the $g$-data. This is important as it
means that at least in the present case the estimates from
$W(\beta_c,L)$ are intrinsically more precise.

It was estimated in \cite{deng:03} that
\begin{equation}
  g(\beta_c,L) = 0.69778(13) \,\left( 1 +
  0.1788(36)\,L^{-0.82(3)}+\cdots\right)
\end{equation}
and our $g$-data, Figure~\ref{fig:6}, are in excellent agreement with this
correction factor for $g(\beta_c,L)$.  
%and plotting this in the same picture as our $g$-data (we did not do
%this {\bf SHOULD WE?}), Figure~\ref{fig:7}, it is hard to tell them
%apart.  For our $L\ge 12$ the relative difference is on the order of
%$\pm 0.002$. 
We estimate the critical values to be $W(\beta_c,\infty) = 0.468(2)$
and $g(\beta_c,\infty)=0.697(2)$, see Figure~\ref{fig:7}, where the
error stems from which points are excluded from the fit.  The value
for $g$ agrees with the formula above but the accuracy is not as good.
Also, we would like to mention that at the temperature where the
magnetisation distribution shifts from unimodal to bimodal, i.e. where
$[\partial^2 p(m)/\partial m^2]_{0}=0$, we found the asymptotic
value of $W$ to be about $0.208$ and for $g$ it takes a value near
$0.433$.

%The extrapolation of the $[W(\beta,L),W(\beta,2L)]$ intersection
%temperatures to infinite size leads to an estimate $\beta_c =
%0.2216539(2)$ {\bf to check}. The equivalent plot for the $g$
%intersections leads to an estimate $\beta_c = 0.22165xx(2)$. 

There are already many accurate estimates of $\nu$ for the $3$d Ising
universality group. Renormalization group studies \cite{guida:97} give
$\nu = 0.6304(13)$ and $\nu = 0.6305(25)$. The main difficulty
concerning either HTSE or MC analyses lies principally with the
problem of properly allowing for corrections to scaling. The
amplitudes of the corrections vary from system to system, favorizing
meta-analyses of data on many systems in the same class.  Butera and
Comi \cite{butera:02} obtain $\nu=0.6299(2)$ from a global analysis of
HTSE data for Ising ferromagnets with spin $S$ running from $1/2$ to
$\infty$ on both sc and bcc lattices, all systems lying in the same
universality class. Their sc $S=1/2$ HTSE results standing alone were
consistent with this value but were less accurate ($0.632(2)$ or
$0.6277(30)$ depending on the analysis method used). Deng and Bl\"ote
\cite{deng:03} obtain an entirely independent global estimate $\nu =
0.63020(12)$ from simultaneous Monte Carlo analyses on a set of eleven
systems all in the same universality class.  It is gratifying that the
present results on one single system are consistent with and
practically as accurate as these global "best estimates" from HTSE and
MC. It would be interesting to establish whether the weak FSS
correction for $[\partial W(\beta,L)/\partial\beta]_{\beta_c}$ is a
general property or is specific to this particular system.

\section{Conclusion}
We introduce an alternative distribution "shape" parameter
$W(\beta,L)$ for numerical studies of the critical properties of model
systems. As an illustration we use this parameter in an analysis of
extensive data sets obtained through a density of states technique
applied to simple cubic $S=1/2$ Ising ferromagnet samples of size up
to $L=256$.  In this system at least, corrections to scaling for
$W(\beta_c,L)$ are considerably weaker than those for the canonical
Binder cumulant $g(\beta_c,L)$ and the equilibration time to obtain
data to a similar degree of precision is significantly lower.  We
obtain estimates for the critical inverse temperature $\beta_c =
0.2216541(2)$ and the critical exponents $\nu = 0.6308(4)$ and
$\omega=0.814(30)$, based only on $W$-data, which are compatible with
and almost as accurate as values from previous Monte Carlo
\cite{deng:03} and high temperature series expansions
\cite{butera:02}.

\section{Acknowledgements}
This research was conducted using the resources of
High Performance Computing Center North (HPC2N).
We would like to thank Paolo Butera for his invaluable advice.

{99}

\end{document}